\documentclass[review]{elsarticle}

\makeatletter
\def\ps@pprintTitle{%
 \let\@oddhead\@empty
 \let\@evenhead\@empty
 \def\@oddfoot{\centerline{\thepage}}%
 \let\@evenfoot\@oddfoot}
\makeatother

\usepackage{lineno,hyperref}
\usepackage{amsmath}
\usepackage[margin=2.5cm]{geometry}

\begin{document}

\begin{frontmatter}

\title{Inferring temporal dynamics from cross-sectional data using Langevin dynamics}





\author[1,2]{Pritha Dutta\corref{contrib}\corref{corauthor}}
\author[3,4]{Rick Quax\corref{contrib}}
\author[3,5]{Loes Crielaard}
\author[2,3,6]{Peter M.A. Sloot}

\cortext[corauthor]{Corresponding author; pritha002@e.ntu.edu.sg}
\cortext[contrib]{Authors contributed equally}

\address[1]{Interdisciplinary Graduate Programme, Nanyang Technological University, Singapore 637335, Singapore.}
\address[2]{Complexity Institute Singapore, Nanyang Technological University, Singapore 637335, Singapore.}
\address[3]{Institute for Advanced Study, University of Amsterdam, Amsterdam 1012 GC, The Netherlands}
\address[4]{Computational Science Lab, University of Amsterdam, Amsterdam 1098 XH, The Netherlands}
\address[5]{Department of Public Health, Amsterdam UMC, University of Amsterdam, Amsterdam Public Health Research Institute, Meibergdreef 9, Amsterdam, The Netherlands.}
\address[6]{HPC Institute, ITMO University, St. Petersburg, The Russian Federation.}

\begin{abstract}
Cross-sectional studies are widely prevalent since they are more feasible to conduct compared to longitudinal studies. However, cross-sectional data lack the temporal information required to study the evolution of the underlying processes. Nevertheless, this is essential to develop predictive computational models which is the first step towards causal modelling. We propose a method for inferring computational models from cross-sectional data using Langevin dynamics. This method can be applied to any system that can be described as effectively following a free energy landscape, such as protein folding, stem cell differentiation and reprogramming, and social systems involving human interaction and social norms. A crucial assumption in our method is that the data-points are gathered from a system in (local) equilibrium. The result is a set of stochastic differential equations which capture the temporal dynamics, by assuming that groups of data-points are subject to the same free energy landscape and amount of noise. Our method is a ‘baseline’ method which initiates the development of computational models which can be iteratively enhanced through the inclusion of expert knowledge. We validate the proposed method against two population-based longitudinal datasets and observe significant predictive power in comparison with random choice algorithms. We also show how the predictive power of our ‘baseline’ model can be enhanced by incorporating domain expert knowledge. Our method addresses an important obstacle for model development in fields dominated by cross-sectional datasets.
\end{abstract}


\end{frontmatter}


\section{Introduction}
Longitudinal studies require a huge investment in terms of time, money, and effort, depending on the system studied. For instance, biological experiment techniques such as sequencing-based assays destroy cells in order to measure certain concentrations; population-based cohort studies in public health, such as HELIUS \citep{snijder2017cohort}, involve asking each of the participants to visit the hospital, measuring various physiological variables, and assessing psychological well-being through interviews and questionnaires. This leads to a relative abundance of cross-sectional datasets in these fields.
However, the price to pay is that cross-sectional data lack the temporal information needed to study the evolution of the underlying processes. This hampers the development of models which can make predictions (predictive models) or even simulate the effects of interventions (causal models) in these fields. Therefore, in order to utilize the abundant cross-sectional data to study the dynamics of system behaviour it is important to design methods to infer the temporal dynamics from these data.

There are several techniques in the literature for estimating pseudo-longitudinal data from cross-sectional data, which include employing distance metrics and graph theoretical operations \citep{peeling2007making, tucker2009pseudotemporal, li2013modelling, magwene2003reconstructing, campbell2018uncovering, gupta2008extracting, dagliati2019inferring, kumar2017understanding}.  They aim to construct realistic trajectories through the feature space by using techniques such as ordering of the data-points and selecting start-points and end-points based on known class labels. For instance, one way to order data-points is by assuming that the label ‘healthy’ precedes the label ‘diseased’. Another way to order biological samples or RNA-seq data is by using their gene expression levels. Of course, these methods rely on the presence of suitable variables in the dataset, as well as the assumptions about how these labels induce an ordering. Our proposed method, on the other hand, infers the temporal dynamics from the distribution of the data-points, and hence, is not dependent on the ordering of the data-points or the presence of order-inducing variables.

In this work, we propose a method for inferring predictive computational models from cross-sectional data using Langevin dynamics \citep{coffey2012langevin}. We reconstruct the free energy landscape, which is a mapping of all possible states of the data-points in the system, from the cross-sectional data by assuming that groups of data-points having similar features follow similar trajectories. We then derive stochastic differential equations based on Langevin dynamics to capture the temporal dynamics, by assuming that groups of data-points are subject to the same free energy landscape and amount of noise. We validate our method against two population-based datasets form the public health domain. These datasets represent systems involving human interaction and social norms. Our assumption is that the use of Langevin dynamics is valid in scenarios where individuals in groups tend to follow norms and adhere to social conventions \citep{chung2016social, smaldino2015social}. This is because these ‘forces’ would lead individuals to move towards the same norm behaviour, making it possible to identify the ‘force field’ that the individuals are following. There are many cross-sectional studies of human behaviours influenced by social norms, such as physical activity, dietary habits, smoking, and alcohol consumption \citep{ball2010healthy, pelletier2014social, eisenberg2003adolescent, litt2011adolescent}. However, cross-sectional data cannot readily be used to develop predictive computational models to study how these behaviours evolve over time. Predictive computational models may however be valuable in this context since they enable the assessment of competing hypotheses by allowing us to evaluate hypothetical scenarios in silico and simulate the effect of interventions. This is especially advantageous for systems for which comparing counterfactual scenarios would not be possible in vivo, as is the case for systems involving human interaction and social norms. We would, for instance, never be able to conduct an empirical study to assess the effect of group-level social norms versus individual-level health-related behaviour on body weight \citep{crielaard2020social}.

The goal of our study is to formulate a method to infer a computational model that predicts the temporal evolution of the data-points. We would like to highlight here that our proposed method is not a causal inference technique in itself, which is a term reserved for automated techniques that identify causes of an effect by establishing that a cause-and-effect relationship exists purely based on data \citep{morgan2015counterfactuals}. Our proposed method selects a stochastic differential equation model which best fits the landscape dynamics derived from the data. Even though a differential equation can always be interpreted as a causal model, since it specifies how one variable changes as function of other variable values, we would like to clarify that our optimally predictive model does not by itself encode valid causal relationships. The selected model by our method (without further constraints) is therefore not causally interpretable. In other words, it can be used to predict the system behaviour, but not how this behaviour will change in response to interventions. To illustrate this point, consider a dataset of body weights of a group of individuals. It may be reasonable to predict that individuals with a weight that is lower than the norm weight will have a statistical tendency to increase their body weight, and vice versa. However, it is an entirely different question what would happen to an individual's body weight after an intervention, for instance, a sudden and large change in diet or physical activity level. Our method does not address the question of `how' mechanistically an individual's body weight increases or decreases, only `that' it will statistically tend to increase or decrease. Our method provides an easy-to-execute tool to infer predictive computational models from these datasets which would enable understanding the progression of these processes.

The causal interpretation of our model can be increased by adding expert knowledge to our presented `baseline' method in the form of (constraints on the) causal relationships between the different variables in the system. This should in turn increase the out-of-sample predictive power (generalizability). For example, in a previous work \citep{crielaard2020social}, we constructed an expert-informed causal model between individual body weight perception, individual weight-related behaviour, and group-level norms towards body weight. We incorporated expert knowledge to gather statements of causal and non-causal relationships to infer a computational model which can be causally interpreted. These statements of causal relations (e.g., `physical exercise directly affects weight loss') essentially constitute constraints on the functional forms of the differential equations that are permitted (the differential equation for `weight loss' must at least include a variable `physical exercise'). The remaining uncertainties (parameter values, functional forms) were then estimated from a cross-sectional dataset by using assumptions. Our proposed method can be considered as a generalization and formalization of this model-building process for systems where groups of individuals adhere to social norms: in its naive form it produces a predictive model, and the more expert-informed constraints on functional forms are added based on causal or non-causal statements, the better the causal interpretation of the resulting model. 

We present the theoretical foundation and a numerical illustration of the method which estimates the dynamics from a cross-sectional data and highlight the assumptions. We validate the technique using two population-based longitudinal datasets where the first time-point is used as the cross-sectional data and the subsequent time-points are used for model validation.

\section{Methods}
\begin{figure*}[t!]
    \centering
    \includegraphics[width=\textwidth]{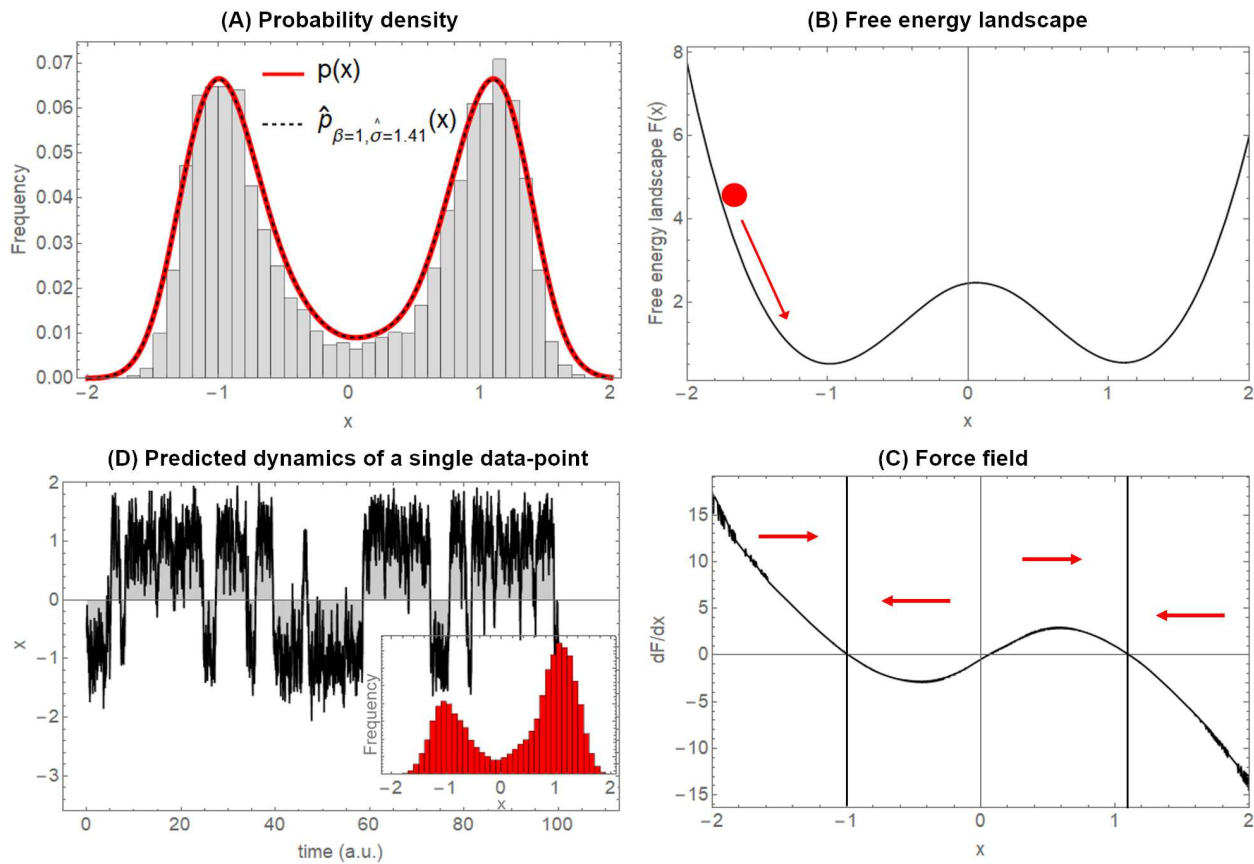}
    \caption{The proposed procedure to infer a computational model from a cross-sectional dataset. (A) The energy landscape $F(x) = -a x^2 + b x^4$ is used to generate a surrogate dataset, using the true probability density (solid red line). The dashed black line is the estimated probability density. (B) From the estimated probability density and the Boltzmann distribution we estimate the free energy landscape $F(x)$. The red ball represents a data-point moving downslope towards the attractor as shown by the red arrow. (C) The force field which is the derivative of the free energy landscape leads to the deterministic dynamics of the Langevin equation. The red arrows show the direction in which a data-point in any of the four partitions will be forced to move by the force field. The regions around 1 and -1 correspond to the two attractors. (D) The resulting predicted dynamics of a single data-point and the occasional transition from one attractor to the other, across the tipping point at $x= 0$. The inset figure is the histogram of the values of this data-point across time. The histogram almost captures the shape of the histogram of the cross-sectional data shown in (A) and is slightly different because we have taken a short time range to improve visibility of the predicted dynamics.}
    \label{fig:free_energy_landscape}
\end{figure*}

\subsection{Langevin dynamics}
We consider that each data-point, denoted as $\vec{x}$, is vector-valued and can change over time. A data-point represents a set of attributes (such as BMI) of an individual. We assume that all data-points in a cross-sectional dataset of size \textit{N} have had sufficient time to mix and explore the state space by the time at which they are observed. Thus, we assume that even if there was a major perturbation, a system of data-points has converged to a stable distribution at the time of our observation. This implies that all the data-points follow a probability distribution, $p(x)$, which is stationary. Further, this implies that \textit{N} is large enough to effectively estimate $p(x)$ from the dataset.

In general it is not possible to derive the temporal evolution, $dx/dt$, from the stationary probability distribution $p(x)$. This is because there are multiple $dx/dt$ which can lead to the same stationary distribution. For example, a data-point rotating clock-wise and another data-point rotating anti-clockwise both have the same circular stationary distribution, but their $dx/dt$ are different. We solve this problem by assuming that each data-point tends to follow the same free energy landscape, $F(x)$, in a `downslope' manner, i.e., moves in time in the direction of $-dF/dx$. $F(x)$ effectively assigns an energy value to each possible vector $\vec{x}$. The assumption is that systems have a tendency to minimize their energy, although random fluctuations may sometimes have the opposite effect. The minimum energy states correspond to the attractors in $F(x)$. This assumption could be valid for social systems where individuals tend to follow norms, adhere to social conventions, and identify themselves with groups \citep{chung2016social, smaldino2015social}. These influences could lead individuals to move towards the same attractors. Thus, in these applications, the probability distribution of data-points could be used to make predictions, whereas in other applications this may not be the case.

A free energy landscape can be considered analogous to an uneven hillside and the data-points in the landscape are the balls rolling down the hillside. The balls will eventually come to rest in a stable state in the valleys. These valleys are the attractors in the free energy landscape. Trajectories across a landscape correspond to the progression of the data-points. Our method estimates the free energy landscape based on the probability distribution of the data-points. Roughly speaking, the free energy landscape is approximately the inverse of the probability distribution of the data-points. Thus, the attractors in this estimated energy landscape will approximately correspond to the peaks in the probability distribution, which correspond to the norms. For instance, a free energy landscape estimated from a probability distribution of the body weights of a group of individuals will have attractors that will correspond approximately to the most probable weights in that group of individuals.

The energy landscape, $F(x)$, can be derived from $p(x)$ through the Boltzmann distribution \citep{swendsen2020introduction},

\begin{equation}
    \beta F(x) = -\log p(x).
    \label{eq:boltzmann}
\end{equation}

Here, the constant $\beta$ is interpreted as the inverse of temperature, or noise level: lower the value of $\beta$, larger is the effect of random fluctuations on \textit{x}, and thus lower the influence of $F$ on the points. The assumption behind this relation is that the data-points are non-interacting. The other assumption, i.e., the data-points are in equilibrium, is already met by the assumption of $p(x)$ having reached stationarity.

In addition to this deterministic tendency given by $F$, there is a random movement (noise) which is irrespective of $F$ and is uncorrelated over time, defined by a Wiener process $W(t)$ \citep{durrett1996stochastic}. Thus, the displacement of each data-point consists of the sum of a deterministic component and a stochastic component which leads to the following overdamped Langevin dynamics equation \citep{coffey2012langevin}:

\begin{equation}
    \frac{dx}{dt} = -\beta \frac{dF}{dx} + \sigma \frac{dW}{dt}.
    \label{eq:langevin}
\end{equation}

Here, $\beta$ controls the relative strength of the deterministic force and $\sigma$ controls the relative strength of the noisy movement. When $\beta \rightarrow 0$, the deterministic component becomes negligible, and thus the data-points diffuse randomly over the state space in all directions. When $\sigma \rightarrow 0$, the stochastic component becomes negligible , and all data-points converge to one or more local minima of $F(x)$, after which no further change occurs. Clearly, a balance is needed between these two opposing processes, which will control the degree of clustering and the amount of variance observed in the predicted distribution, $\hat{p}_{\beta,\sigma}(x)$, and match with the $p(x)$ from the data.

For determining the values of $\beta$ and $\sigma$, it is important to realize that only their ratio changes the stationary data distribution. Thus, if we fix the timescale of the deterministic component by setting $\beta=1$ without loss of generality, then the remaining free parameter $\sigma$ controls the ratio of the random and deterministic forces. This timescale can be fixed because if we multiply both parameters in Eq.~\ref{eq:langevin} with a constant $\tau$, the velocity $dx/dt$ is also multiplied by $\tau$. This means that $x$ changes $\tau$ times faster over time. In principal, it is impossible to derive how fast $x$ changes per unit time from the data available at a single time-point. We can, however, still predict the directions of the data-points, which is our main interest.

\subsection{Numerical algorithm}
To explain the numerical algorithm we will generate a set of data-points moving over a known free energy landscape and show how the algorithm recovers the dynamics. Consider a true underlying free energy landscape, $F(x) = -a x^2 + b x^4$, from Landau's second order phase transition formalism, which contains two attractor states as long as $a>0, b>0$. We set $a=3$ and $b=1$, where $a$ controls the height of the energy barrier separating the two attractors. Assuming the Boltzmann distribution (Eq.~\ref{eq:boltzmann}), we generate a dataset consisting of 5000 i.i.d. data-points from the probability density function, $p(x)=e^{a x^2 - b x^4}/Z$, where $Z$ is the normalizing constant. $p(x)$ is shown in Fig.~\ref{fig:free_energy_landscape}A as a solid red line. The sampling is done through the inverse transform sampling method \citep{gentle2006random}.

We will now treat this dataset as given and `forget' the $F(x)$ used to generate it. The true $p(x)$ always has an exponential form due to the assumption of a Boltzmann distribution. Therefore, a Gaussian kernel density estimation algorithm is used to estimate the data distribution. A Gaussian kernel density estimate is defined as $\hat{p}(x) = (nh\sqrt{2\pi})^{-1} \sum_i^n e^\frac{(x-x_i)^2}{2h^2}$, where $x_i$ are the data-points and the parameter $h$ is determined using Silverman's optimization procedure \citep{silverman1986density}. We would like to highlight here that even though the parameters $h$ and $\sigma$ represent similar concepts (both are standard deviation controlling parameters), they implement different components. The parameter $h$ controls the standard deviation of the Gaussian kernel density estimate obtained from the data, whereas, $\sigma$ controls the the ratio of the random and deterministic forces in Eq.~\ref{eq:langevin}. The estimated $\hat{p}(x)$ from the dataset is shown in Fig.~\ref{fig:free_energy_landscape}A as a dashed black line. This estimated $\hat{p}(x)$ is then used to estimate the free energy landscape as $\hat{F}(x) = -\log{\hat{p}(x)}$. By fixing the timescale of the deterministic dynamics through $\beta=1$, the deterministic term, $dF/dx$, of the Langevin dynamics (Eq.~\ref{eq:langevin}) is given as,

\begin{equation}
    \frac{dF}{dx} = -\sum_{i=1}^n \frac{\left(\frac{x_i e^{-\frac{\left(x-x_i\right){}^2}{2 h^2}}}{h^2}-\frac{x e^{-\frac{\left(x-x_i\right){}^2}{2 h^2}}}{h^2}\right)}{e^{-\frac{\left(x-x_i\right){}^2}{2 h^2}}}.
\end{equation}

The stochastic part is then fully determined by an optimal choice for $\sigma$. The optimal $\sigma$ is determined by using the Hellinger distance, $H\left( \hat{p}_{\beta=1,\sigma}(x), p(x) \right)$, as the cost function. The Hellinger distance is defined as, $H(p,q) = \frac{1}{2} \int_x \left( \sqrt{dp(x)} - \sqrt{dq(x)} \right)^2 dx$. To evaluate the goodness of fit for each choice of $\sigma$ during this optimization process, we need to estimate the stationary probability density $\hat{p}_{\beta=1,\sigma}(x)$ numerically. Since integration techniques for stochastic differential equations are computationally intensive, we choose to discretize the domain of the data, $x$, into discrete points with distance $\Delta{x}$, which is calculated as $\Delta{x} = \sqrt{\Delta{t}}/f$.
We use time-step $\Delta{t}$ such that $\sigma \sqrt{1000 \Delta{t}}=(x_{\text{max}}-x_{\text{min}})/2$. This means that it would take a random Wiener diffusion process with diffusion parameter $\sigma$ about 1000 time-steps to reach the entire range of $x$ values. The expected displacement of a random diffusion process after $\Delta{t}$ time is $\sqrt{\Delta{t}}$, which we divide up into $f$ discrete points. Thus, $f$ controls the `fineness' of this grid. We find that $f=10$ produces accurate results at low memory cost. $x_{\text{min}}$ and $x_{\text{max}}$ are calculated from the data by taking the floor of the minimum value and ceiling of the maximum value from the standardized data respectively. In this case, we get $x_{\text{max}}=-x_{\text{min}}=3$ from the generated dataset. We obtain 601 discrete values for $x$ and a sparse, approximately band-diagonal transition matrix of dimension $601 \times 601$, denoted by $W$. To calculate the transition probabilities in $W$, we first determine the possible positions that a data-point, starting at position $x$, can reach after $\Delta{t}$ time-steps. The displacement due to the deterministic force is given by $x+(-\frac{dF}{dx} \Delta{t})$. Taking this value as the mean, we determine from the 601 discrete values of $x$ the values ($y$) that fall within 4 standard deviations of this mean, i.e., within $\pm 4\sigma \sqrt{\Delta{t}}$ of the mean. We then calculate the probabilities of displacement to these $y$ values due to random diffusion, by assuming a normal distribution with mean as $x+(-\frac{dF}{dx} \Delta{t})$ and standard deviation as $\sigma \sqrt{\Delta{t}}$. Thus, $W_{x \xrightarrow y} = \mathcal{P}\left(y \mid \mathcal{N}\left(x-\frac{dF}{dx} \Delta{t}, \sigma \sqrt{\Delta{t}}\right) \right)$, where $\mathcal{P}()$ denotes probability density function.

For discrete Markov processes, the stationary distribution vector $\pi$ can be found directly by solving the set of linear equations $\pi = W \pi$, which is computationally an efficient operation since it reduces to finding the (left) eigenvector of $W$, having an eigenvalue of 1. Before performing this operation, we normalize the rows of $W$. The initial distribution vector, $\pi_0$, is calculated from the data as, $\pi_0(x) = \int_{x-\Delta{x}/2}^{x+\Delta{x}/2} p(x)$. Finally, we compute the Hellinger distance, $H\left(\hat{p}_{\beta=1,\sigma}(x), p(x)\right)$, for which $\pi$ is converted to a continuous function, $\hat{p}_{\beta=1,\sigma}(x)$, using third-order spline interpolation. If changing $\sigma$ no longer reduces the Hellinger distance, the procedure terminates and the Langevin dynamics (Eq.~\ref{eq:langevin}) is completely specified. In Fig.~\ref{fig:free_energy_landscape}D the resulting predicted dynamics of a single data-point is shown, where the optimal $\hat{\sigma} \approx 1.41$.

An interesting consequence of having this dynamical model is the ability to quantify the relative frequency of tipping point transitions of data-points as well as the controllability of the tipping point transitions. This frequency is relative because, as mentioned before, there is no absolute time scale; it must be compared to, for example, the time scale of the short (within-attractor) fluctuations. Since short fluctuations can often be measured experimentally or assigned an approximate time scale, this leads to the prediction of how long it takes a data-point to transit between attractors. Additionally, it provides a prediction of how much `effort' (or energy) would be needed to make transitions occur faster (if desirable, such as towards an attractor representing a healthier body mass index (BMI)) or slower (if the BMI represented by the attractor is already healthy). For instance, an increase in $\sigma$ pertains to adding more random perturbations making transitions more frequent in both directions. Another interesting possibility is to add a term to the free energy function pointing towards the desired attractor state, inducing an asymmetric force.

Suppose, we introduce an asymmetric intervention by adding a term to the free energy function which makes the left attractor preferred. The free energy function is then represented as, $G(x)=F(x)+cx$. Here, $c$ controls the strength of the effect of the intervention. To determine the relative amount of effort required for such an intervention, we consider the relative change of velocity of an individual due to this intervention, denoted by $r(c)$. $r(c)$ is defined as the square root of the second moment of displacement. We assume that $W_2$ denotes a Wiener process at time t with mean $\frac{dG}{dx}t$ and standard deviation $\sigma \sqrt{t}$, and $W_1$ denotes a Wiener process at time t with mean $\frac{dF}{dx}t$ and standard deviation $\sigma \sqrt{t}$. The second moments of the above two Wiener processes are denoted by $E[W_2^2]$ and $E[W_1^2]$ respectively. Thus, $r^2$ is defined as,

\begin{eqnarray}
    r^2 &=& \int_{-\infty}^{\infty} p(x) \frac{E[W_2^2]-E[W_1^2]}{E[W_1^2]} \, dx, \nonumber \\
        &=& \int_{-\infty}^{\infty} p(x) \frac{t \left((dG/dx)^2-(dF/dx)^2\right)}{(dF/dx)^2 t+\sigma ^2} \, dx, \nonumber \\
        &=& \int_{-\infty }^{\infty } p(x) \frac{t \left(\left(2ax-4bx^3-c\right)^2-\left(2ax-4bx^3\right)^2\right)}{t \left(2ax-4bx^3\right)^2+\sigma ^2} \, dx, \nonumber \\
        &=& \int_{-\infty }^{\infty } \frac{p(x)t \left(-4axc+8bx^3c+c^2\right)}{t \left(2ax-4bx^3\right)^2+\sigma ^2} \, dx, \nonumber \\
        &=& \int_{-\infty }^{\infty } \frac{p(x)t \left(-4axc+8bx^3c+c^2\right)}{t \left(2ax-4bx^3\right)^2+\sigma ^2}\times \frac{t^{-1}}{t^{-1}} \, dx, \nonumber \\
        &=& \int_{-\infty }^{\infty } \frac{p(x) \left(-4axc+8bx^3c+c^2\right)}{\left(2ax-4bx^3\right)^2+ \frac{\sigma ^2}{t}} \, dx.
        \label{eq:rc1}
\end{eqnarray}

Here, $p(x)=e^{a x^2 - b x^4}/Z$, where $Z$ is the normalizing constant. From Eq.~\ref{eq:rc1}, we see that $r\propto \frac{\sqrt{t}}{\sigma}$. Thus, $r$ depends more strongly on $\sigma$ ($r \propto \sigma^{-1}$) than it does on $t$ ($r \propto \sqrt{t}$), since the power of $t$ is half that of $\sigma$. To set $t$ to an interpretable value, we choose the mean time between tipping point (at $x=0$) transitions, estimated by integrating the Langevin dynamics equation (Eq.~\ref{eq:langevin}) over time between 0 to $10^6$. A tipping point transition is observed when two consecutive values of $x$ have different signs. Using this method, we obtain $t_c \approx 0.0013$ in the (arbitrary) time units of the process. Filling in $t_c$, $\hat{\sigma}$, and $c=1$ as an example, we find that $r \approx 0.025$. This means that, compared to all the existing (deterministic and random) forces already applying to $x$ such as social norms (summarized into $F$ and $\sigma$), $2.5\%$ additional force would be required in order to make the left-hand attractor contain about $\int_{-\infty}^0 e^{-G} dx / \int_{-\infty}^\infty e^{-G} dx \approx 85.1\%$ of the individuals in equilibrium, instead of $50\%$ under the symmetric $F$. We note that it is impossible to say in absolute terms how much effort is needed to implement the intervention, which would entail listing all existing current forces that lead to the currently observed distribution. We also note that this cannot be used to infer \emph{how} an intervention should be implemented, i.e., which are the causal relationships which should be exploited in order to achieve such an intervention (e.g., stimulating physical activity versus healthier diet).

\subsection{Validation procedure}
The displacement of an individual \textit{i} for a $\Delta t$ time-increment is represented by a normal distribution with the displacement due to the deterministic force, $(\frac{-dF}{dx}\Delta t)^i$ as the mean and $\sigma\sqrt{\Delta t}$ as the standard deviation. Using this normal distribution, we determine for each individual \textit{i} the probability of positive displacement and negative displacement, denoted as $P^i_{\text{PD}}$ and $P^i_{\text{ND}}$, respectively.

Here, we quantify the prediction accuracy of our model for individual \textit{i}, which we denote as $A^i$. If the observed displacement for individual \textit{i} from the data is positive, we assign $P^i_{\text{PD}}$ as the prediction accuracy of our model, and if the observed displacement is negative, we assign $P^i_{\text{ND}}$ as the prediction accuracy of our model for individual \textit{i}. We also determine the maximum prediction accuracy of our model for individual \textit{i}, denoted as $A^i_{\text{maximum}}$, as the maximum of the two probabilities $P^i_{\text{PD}}$ and $P^i_{\text{ND}}$. We standardise the data to have zero median and unit standard deviation in order to have a mathematically convenient representation of the variable.

To determine the optimal value of $\Delta$t, we employ the metric of Euclidean distance, by varying its value from 1 to 100 with increments of 1, and selecting the value that gives the minimum Euclidean distance between the displacement due to the deterministic force ($-\frac{dF}{dx}\Delta t$) predicted by our model and the observed displacement from data.

As a further validation step, we define an ideal case based on the expectation that individuals should move towards the model's attractor. Assuming this behaviour, we determine the ‘maximally achievable’ prediction accuracy of our model. The purpose of comparing our model prediction with this ideal case is to test if the prediction accuracy of our model increases when the displacement of individuals is according to the expectation. The displacement in the ideal case is taken as the negative of the standardised data.

In addition to the above validation tests, we compare the prediction accuracy of our model to random choice under the null hypothesis of random displacement. We randomly choose between $P^i_{\text{PD}}$ and $P^i_{\text{ND}}$ for each individual \textit{i} and then take the average over all individuals to obtain the prediction accuracy by random choice. We repeat this random choice step 1000 times to obtain a distribution of mean prediction accuracies of random choice and then determine the 95\% confidence interval of this distribution. If the average prediction accuracy of our model is greater than the upper limit of this confidence interval, which we denote as ${U}_{\text{CI}}$, we can say that our model prediction is significantly better than the prediction obtained by random choice. We also scale the average prediction accuracy (${A}^{\text{average}}$) to better compare with $U_{\text{CI}}$ and the average maximum prediction accuracy (${A}^{\text{average}}_{\text{maximum}}$) as

\begin{equation} \label{eq:val1} 
{A}^{average}_{scaled}=\frac{{A}^{average}-{U}_{CI}}{{A}^{average}_{maximum}-{U}_{CI}} 
\end{equation} 

After the above scaling, ${U}_{\text{CI}}$ will correspond to zero and ${A}^{\text{average}}_{\text{maximum}}$ will correspond to 1. If ${A}^{\text{average}}_{\text{scaled}}$ is above zero, we say that our model prediction is better than random choice; if it is below zero, then our model prediction is indistinguishable from random choice. Thus, we validate our model predictions against data as well as test whether the prediction accuracy of our model is significantly better than that by random choice.

\subsection{Data}
We validate our model against two population-based longitudinal datasets: the College study dataset \citep{lowe2015short} and the Hoorn study dataset \citep{van2010relationship, koopman2017association, rutters2018cohort}, where we consider the first time-point as the cross-sectional data and the subsequent time-points are compared to the model prediction.

The College study dataset \citep{lowe2015short} is a longitudinal dataset with weight measured over 5 time-points: baseline, 6 weeks, 6 months, 12 months, and 24 months. 
This study was conducted with 294 female first-year students [age: $18.24 \pm 0.44$ years; all values are expressed as mean $\pm$ SD unless otherwise specified] recruited from two universities in Philadelphia. We preprocessed this dataset to include only those participants who had their weights measured for all the 5 time-points and obtained 162 participants. We calculated the BMI (in $\mathrm{kg/m^2}$) of these 162 participants from their weights and heights and their baseline BMI distribution is $23.59 \pm 2.69$ $\mathrm{kg/m^2}$.

The Hoorn study dataset \citep{van2010relationship, koopman2017association, rutters2018cohort} is a longitudinal dataset with BMIs measured over 2 time-points: baseline, and 7 years. The baseline study \citep{van2010relationship, rutters2018cohort} was conducted in 2006-2007 in the Dutch city of Hoorn and included 2,807 participants. The follow-up study \citep{koopman2017association, rutters2018cohort} was conducted in 2013-2015 which included 1,734 participants out of the 2,807 participants in the baseline study. We preprocessed this dataset to include only those participants who had their BMI measured for both the time-points and obtained 1,727 participants with age $53.62 \pm 6.53$ years and baseline BMI $26.11 \pm 3.88$ $\mathrm{kg/m^2}$.

\subsection{Data and Code availability}
The data that support the findings of this study are available from second party. Restrictions apply to the availability of these data, which were used under license for this study. Data are available from Lowe et al. \citep{lowe2015short} and Rutters et al. \citep{rutters2018cohort} on request.

The code for the proposed method is written in Mathematica programming language and is available at \url{https://github.com/Pritha17/langevin-crosssectional}.

\section{Results}

\begin{figure*}[t!]
    \centering
    \includegraphics[width=\textwidth]{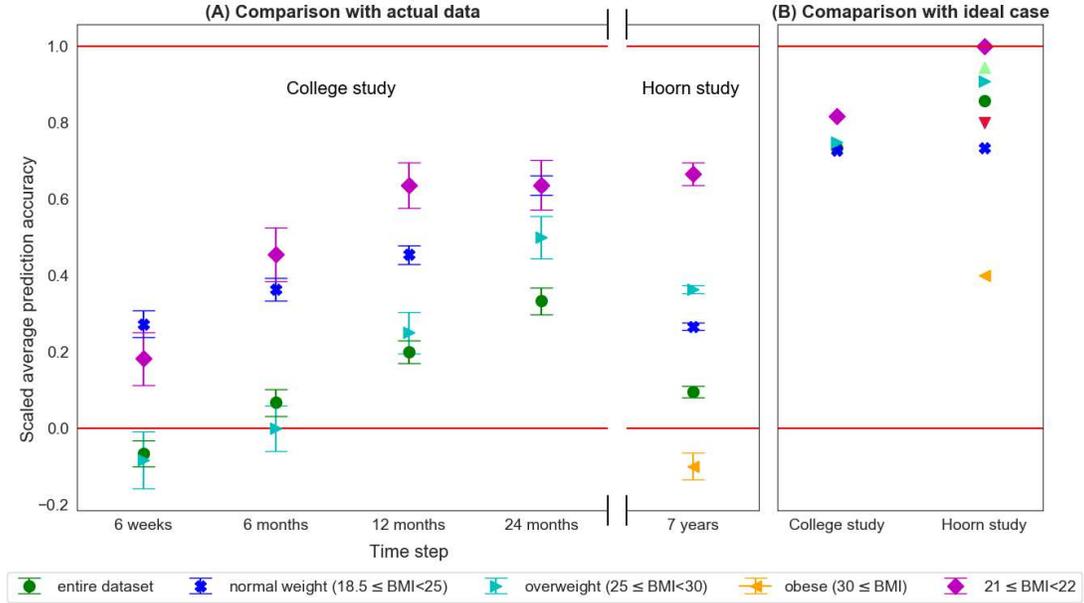}
    \caption{(A) The scaled average prediction accuracies, $\mathrm{{A}_{\text{scaled}}^{\text{average}}}$ (Eq.~\ref{eq:val1}) corresponding to the 4 time-points: 6 weeks, 6 months, 12 months, and 24 months of the College study dataset, and the single time-point of 7 years of the Hoorn study dataset. The vertical bars at each point represent the error bar for the respective model prediction accuracy calculated from 1000 bootstrap samples. The solid red line at 0 corresponds to the upper limit of the 95\% confidence interval of the distribution of mean prediction accuracies of random choice, $\mathrm{{U}_{\text{CI}}}$ (refer to \textbf{Methods: Validation procedure}), and the solid red line at 1 corresponds to the average maximum model prediction accuracy, $\mathrm{{A}^{\text{average}}_{\text{maximum}}}$ (refer to \textbf{Methods: Validation procedure}). Model prediction accuracies above the line at 0 are significantly better than the prediction accuracy of random choice. (B) The ‘maximally achievable’ scaled average prediction accuracy of our model obtained by validating against the ideal case when individuals change their BMI in the expected manner, i.e., decrease or increase their weight tending towards the attractor BMI (refer to \textbf{Methods: Validation procedure}).}
    \label{fig:prediction_accuracy}
\end{figure*}

We present a baseline method of inferring a predictive computational model from cross-sectional data based on assumptions that do not depend on expert knowledge. We validate our model predictions against two longitudinal datasets: the College study dataset \citep{lowe2015short} and the Hoorn study dataset \citep{van2010relationship, koopman2017association, rutters2018cohort}, where we consider the first time-point as the cross-sectional data and the subsequent time-points are compared to the model predictions. The College study dataset contains BMI values over 5 time-points: baseline, 6 weeks, 6 months, 12 months, and 24 months. The Hoorn study dataset contains BMI values over 2 time-points: baseline, and 7 years. Fig.~\ref{fig:prediction_accuracy} shows our model prediction accuracies ($\mathrm{{A}^{average}_{scaled}}$ (Eq.~\ref{eq:val1})) using these two datasets. Our model shows significant predictive power (green circles in Fig.~\ref{fig:prediction_accuracy}(A)) compared to the prediction of a random choice algorithm (solid red line 0 in Fig.~\ref{fig:prediction_accuracy}). Additionally, our model prediction accuracy improves further when we incorporate expert knowledge to our model, as shown by the blue crosses, cyan left triangles, and magenta diamonds in Fig.~\ref{fig:prediction_accuracy}(A).

To test if the predictive power of our model is enhanced with the incorporation of domain expert knowledge, we apply the empirical observation from epidemiology that individuals from different BMI categories follow different landscapes \citep{batterham2016baseline, ostbye2011body}. In accordance, we cluster the datasets based on the standard BMI categories: underweight (BMI $\mathrm{<}$ 18.5), normal weight (18.5 $\mathrm{\le}$ BMI $\mathrm{<}$ 25), overweight (25 $\mathrm{\le}$ BMI $\mathrm{<}$ 30), and obese (30 $\mathrm{\le}$ BMI) \citep{world2000obesity}. We respectively obtain clusters of sizes 0, 119, 39, and 4 from the College study dataset, and 12, 739, 729, and 247 from the Hoorn study dataset. Since the underweight and obese clusters from the College study dataset have only 0 and 4 individuals, respectively, and the underweight cluster from the Hoorn study dataset has only 12 individuals, we disregard those clusters. We observe that if we consider separate attractor landscapes for these different clusters, the model prediction accuracy increases significantly (Fig.~\ref{fig:prediction_accuracy}A). The attractor in each cluster approximately corresponds to the BMI that is most prevalent relative to the group of individuals in that cluster. In addition, we select a narrow BMI range to see if clustering individuals having almost exact BMIs further improves the prediction accuracy. Accordingly, we select all individuals having BMI in the narrow range of 21 $\mathrm{\le}$ BMI $\mathrm{<}$ 22, and obtain 29 individuals from the College study dataset, and 96 individuals from the Hoorn study dataset. We observe that the model prediction accuracy increases further (Fig.~\ref{fig:prediction_accuracy}A). Thus, from the above results we can conclude that the incorporation of domain expert knowledge to our baseline method further enhances the predictive power of our model.


\begin{figure*}[t!]
    \centering
    \includegraphics[width=\textwidth]{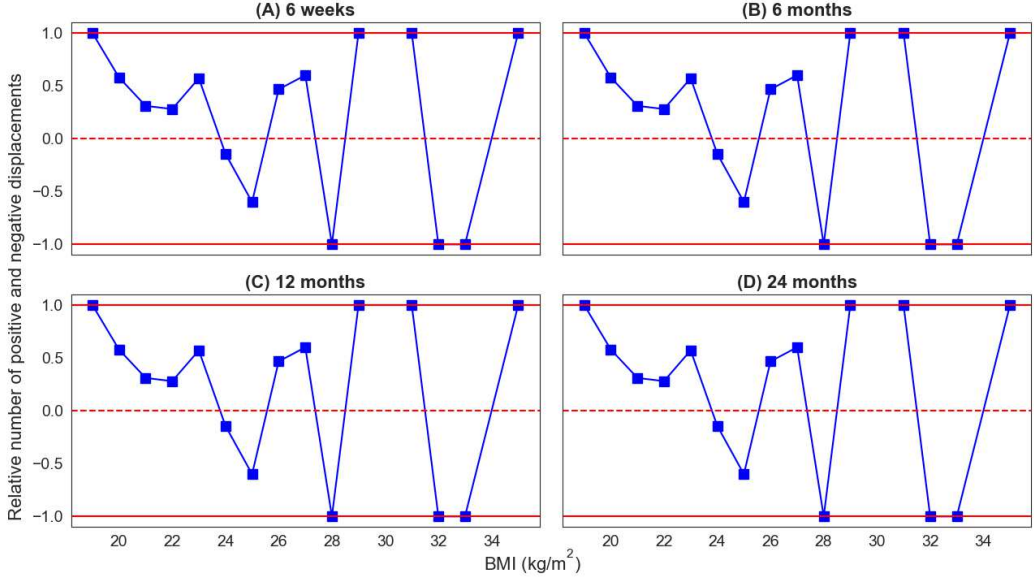}
    \caption{The relative number of positive and negative displacements in each BMI bin at (A) 6 weeks, (B) 6 months, (C) 12 months, and (D) 24 months obtained from the College study dataset. The relative number of positive and negative displacements in each BMI bin is calculated as $\frac{\textrm{number of positive displacements - number of negative displacements}}{\textrm{number of positive displacements + number of negative displacements}}$. This relative number will be 1 or -1 if all displacements at a particular BMI bin are in the same direction (either positive or negative). If a particular BMI bin has mixed displacement directions then the relative number will be between -1 and 1. A positive relative number indicates that there are more positive displacements than negative displacements and vice-versa.}
    \label{fig:displacement_histogram}
\end{figure*}

We also define an ideal case based on the hypothesis that all individuals should move towards the attractor, which corresponds to the norm BMI (the BMI that is most prevalent relative to the group of individuals under consideration). This means that individuals with a BMI that is greater than the attractor BMI should decrease their BMI and vice-versa. Assuming this behaviour, we define a `maximally achievable' model prediction accuracy. As observed from Fig.~\ref{fig:prediction_accuracy}B, these `maximally achievable' model prediction accuracies are significantly higher than the actual prediction accuracies using the real data. This reflects that not all individuals in the dataset are changing their BMI as expected, i.e., individuals with a BMI that is greater than the attractor BMI sometimes increase their BMI and vice-versa. Based on this observation, we further analyse the data to see if individuals having the same BMI indeed have displacements in the same direction. We select 15 BMI bins based on the data (the bins are shown as x axis labels in Fig.~\ref{fig:displacement_histogram}). We place an individual in BMI bin \textit{x} if the individual's BMI falls in the range of $\mathrm{x\le \textrm{BMI}<x+1}$. For example, we place an individual in BMI bin 25 if the individual's BMI falls in the range of $\mathrm{25\le \textrm{BMI}<26}$. Then, we calculate the displacements from baseline to 6 weeks, 6 months, 12 months, and 24 months. Fig.~\ref{fig:displacement_histogram} shows the relative number of individuals having positive and negative displacements in each BMI bin, which is calculated as $\frac{\textrm{number of positive displacements - number of negative displacements}}{\textrm{number of positive displacements + number of negative displacements}}$. If all individuals in a particular BMI bin have displacements in the same direction (positive or negative), then this relative number will be 1 or -1 as shown by the red solid lines in Fig.~\ref{fig:displacement_histogram}. If individuals in a particular BMI bin have mixed displacement directions then this relative number will be between 1 and -1. A positive relative number indicates that there are more positive displacements than negative displacements and vice-versa. We observe from Fig.~\ref{fig:displacement_histogram} that this relative number is between 1 and -1 for most of the BMI bins. Thus, individuals having similar BMIs have both positive and negative displacements indicating that individual behaviour is inherently random.

\section{Discussion}
Cross-sectional studies are widely prevalent since they require less investment in terms of time, money and effort compared to longitudinal studies. However, since these data lack temporal information, they cannot be used directly to study the evolution of the underlying processes. Nevertheless, this temporal information is essential to develop predictive computational models which is the first step towards causal modelling. In this work, we present a novel method to infer predictive computational models from cross-sectional data using Langevin dynamics. This method can be applied to any system that can be described as effectively following a free energy landscape, such as protein folding, stem cell differentiation and reprogramming, and social systems involving human interaction and social norms.

Our method in itself infers predictive computational models from cross-sectional data based on the Langevin dynamics and two assumptions. The first assumption is that of well-mixedness or ergodicity, i.e., the data-points are sufficiently mixed at the time of observation and are at (local equilibrium). Thus, we assume that even if there was a major perturbation, a system of data-points has converged to a stable distribution at the time of our observation. This means that the mean value of all data-points at a particular time-point and the  mean value of any one data-point across time remain the same. This assumption would be valid for short-term where we do not expect any major perturbation to the system. The second assumption is that each data-point tends to follow the same free energy landscape in a `downslope' manner, i.e., two data-points will not follow each other even if they have the same stationary distribution. For instance, one data-point rotating clock-wise and another data-point rotating anti-clockwise will have the same circular stationary distribution, but the changes over time, $dx/dt$, are different. The second assumption is based on the fact that systems have a tendency to minimize their energy and hence, even though the changes over time for the two data-points may be different, they move in time in the same direction, i.e., downslope of a convex free energy landscape. For instance, two individuals will have different rates of change of BMI over time, but both will tend towards the norm BMI.

As opposed to black-box machine learning techniques, our technique is based on interpretable assumptions such that domain expert knowledge can be readily incorporated. That is, the causal interpretation of the resulting model can be enhanced by adding domain expert knowledge in the form of statements of causal and non-causal relationships. This should lead to increased and more robust predictive power, as is indeed demonstrated by our clustering of different categories of BMI.
 
It is important to realize that the proposed method can only estimate directions of progression, not velocities. This is because, in principal, it is impossible to derive how fast a data-point (for instance BMI) changes per unit of time from a single time-point data. In other words, the timescale of the predicted dynamics remains unknown. We can only estimate, for example, whether an individual will increase or decrease his/her BMI, and whether another individual progresses slower or faster. In some cases it is possible to estimate a timescale from the data, for instance by quantifying the relative frequency of tipping point transitions in the model and comparing it to knowledge or data about it. It can also be inferred from known statistical properties of the rates of change in reality; for instance, the fact that the maximum sustainable rate of weight loss observed in a population is about 2 kg per month used in our previous work \citep{crielaard2020social}.
 
The usefulness of our method lies in the fact that it can help to reveal the short-term dynamics from a single time-point, without any dependence on expert knowledge, and this can be used as a starting point for developing causal models with the help of expert knowledge. We believe that the proposed method is sufficiently simple to use as well as interpretable to initiate the iterative development of computational models for any system that can be described as effectively following a free energy landscape and thus help in studying the progression of important processes.

\section{Conclusions}
We have proposed a method to infer predictive computational models from cross-sectional data based on Langevin dynamics for systems that effectively follow a free energy landscape. Our method shows significant predictive power when validated against two independent population-based longitudinal datasets which represent systems involving human interaction and social norms. The performance of our method could be even further improved by taking expert domain knowledge into account. This suggests that our method can indeed be an effective modelling paradigm for systems that can be interpreted as effectively following a free energy landscape. Our method can bootstrap the use of the already abundant cross-sectional datasets to study the evolution of the underlying processes and initiate the iterative development of predictive computational models.

\section*{Author contributions}

R.Q. conceptualised the study and performed the mathematical derivations. P.D. carried out the simulations and analysis. P.D. and R.Q. drafted the manuscript with critical input from L.C. and P.M.A.S. All authors approved the final version.

\section*{Declaration of interests}
The authors declare no competing interests.

\section*{Acknowledgment}

We thank Michael Lowe, Professor (Department of Psychology, Drexel University), for providing us with the College study dataset. We also thank Jeroen Bruggeman, Associate Professor (Department of Social and Behavioural Sciences, University of Amsterdam), Debraj Roy, Assistant Professor (Computaional Science Lab, University of Amsterdam), and Nadege Merabet, PhD (University of Amsterdam) for reviewing the paper.
This work is supported by the NTU Research Scholarship, DINAMICS (ZonMw, Netherlands Organization for Health Research and Development, project number: 531003015), Social HealthGames (NWO, the Dutch Science Foundation, project number: 645.003.002), Computational Modelling of Criminal Networks and Value Chains (Nationale Politie, project number: 2454972), and TO\_AITION (EU Horizon 2020 programme, call: H2020-SC1-2018-2020 , grant number: 848146).

\bibliography{mybibfile}

\end{document}